# An X-ray Extremely wide-field Spectroscopic Surveyor (XESS, or “chess”)[1]

**Authors:** W. Peter Maksym (NASA MSFC, walter.p.maksym@nasa.gov)[2], Melville Ulmer (Northwestern University), Stephan Friedrich (LLNL), Francesco Tombesi (INAF Rome), Riccardo Arcodia (Harvard University), S. Panini Singam (NASA MSFC/USRA), Stephen Bongiorno (NASA MSFC), Nicholas Thomas (NASA MSFC), Martin Elvis (Smithsonian Astrophysical Observatory), Ming Sun (University of Alabama, Huntsville), John ZuHone (Smithsonian Astrophysical Observatory), M. Lynne Saade (NASA MSFC/University of Alabama, Huntsville), Philip Kaaret (NASA MSFC), Doug Swartz (NASA MSFC/USRA), Erin Hicks (University of Alaska), Ming-Yi Lin (University of Toledo), Pallavi Patil (NRAO), Jimmy Irwin (University of Alabama, Tuscaloosa), Shrabani Kumar (NASA MSFC/USRA), Kirtan Dixit (NASA MSFC/USRA), Sudip Chakraborty (NASA MSFC/USRA), Andrea Gnarini (NASA MSFC/ORAU), Steven Ehlert (NASA MSFC)

**A White Paper in Reponse to the Ad ASTRA Initiative.**
For ExoPAG, CoPAG, PhysPAG

## 1. Science Investigation:

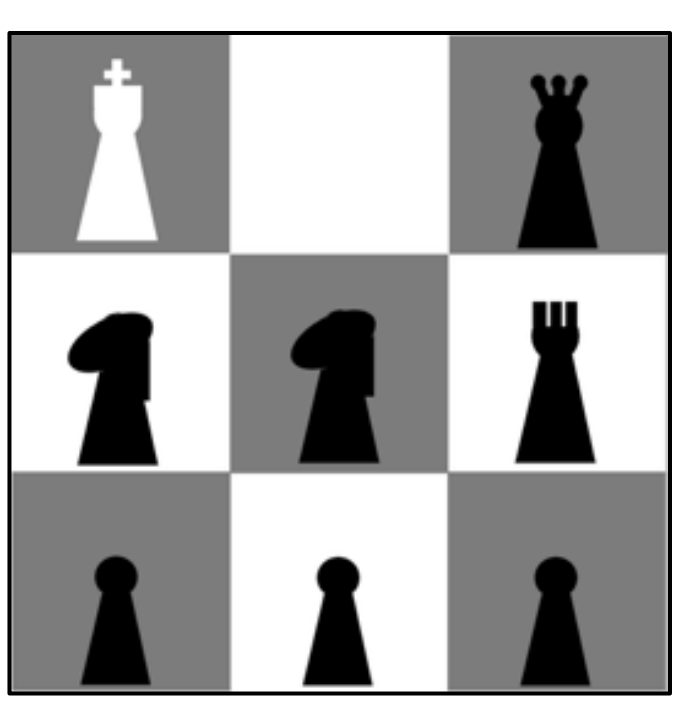

We propose an X-ray observatory, the X-ray Extremely wide-field Spectroscopic Surveyor (XESS, pronounced “Chess”) to ***bring high energy resolution X-ray spectroscopy into the era of synoptic surveys and of time domain and multi-messenger astrophysics*** (TDAMM). Such an observatory is optimally suited to addressing key elements of the 2020 Decadal Survey on Astronomy and Astrophysics.

“Pathways to Discovery in Astronomy and Astrophysics for the 2020s” lays out a comprehensive approach to many scientific goals best served by X-ray observations [1]. These goals include understanding black hole (BH) mass and spin distributions [2,3], BH formation and evolution [4] and the equation of state of ultra-dense matter [5]. We want

[1] Artificial Intelligence (AI) Usage Disclosure: This document was created with assistance from Microsoft Copilot, which was used to search literature and inform concept choices, to develop preliminary estimates of cost, and to augment optics trade-off analysis. The content has been reviewed and edited by W. P. Maksym. More information on the extent and nature of AI usage is included in the Acknowledgements section of the document.

[2] Corresponding author email address

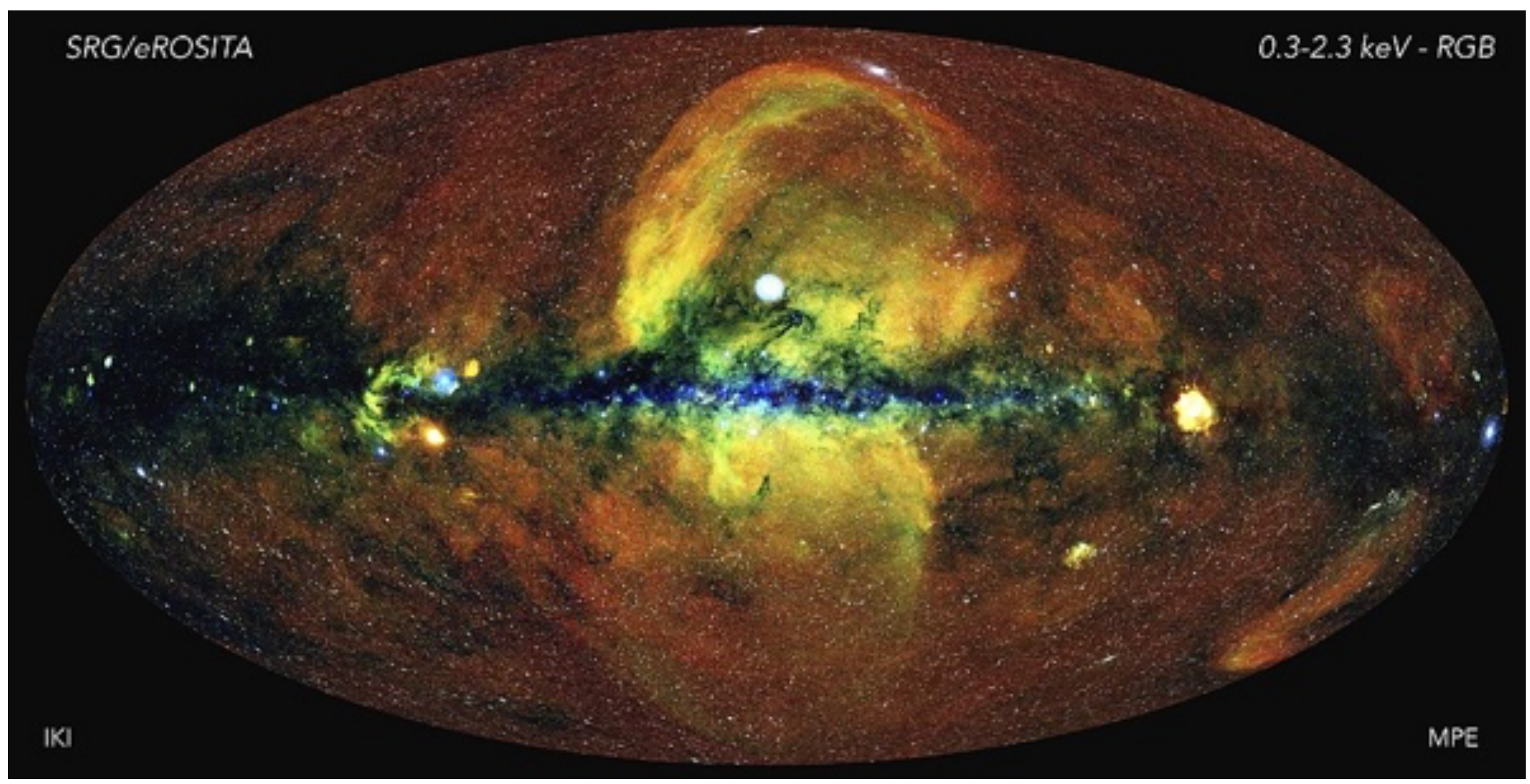

*Figure 1: The eROSITA All Sky Survey (X-rays). XESS will provide order-of-magnitude increases in sensitivity and spectroscopic resolution, and will greatly expand beyond its time domain capabilities. Image: J. Buchner, CC-BY-SA 4.0*

to know about when and how transients are powered by neutron stars (NSs) or BHs, the role of shocks, and novel, unexplored transient phenomena [6,7]. We want to know how the seeds of BHs form and grow, and how electromagnetic phenomena can coordinate with gravitational waves (GWs) and particle-driven signatures (e.g. neutrinos) [4,8,9]. We aim to understand the expansion of the universe, the formation of large-scale structure, and the nature of dark matter and dark energy [10]. We need to understand the co-evolution of galaxies and BHs, the role of feedback, the flow of hot plasmas on scales from the event horizon to galaxy cluster outskirts, the chemical evolution of the universe, dust and gas [11,12,13,14]. And we must know the evolution of these phenomena over cosmic time.

The role of sensitive, high energy resolution X-ray spectroscopy for answering these questions is so far-reaching that we cannot hope to cover them all within a single white . The main gap we propose to address, however, is critical and strategic.

*eROSITA* has made clear the power of X-ray synoptic surveys [15]. *XRISM* points to the power of sensitive X-ray spectroscopy [16], and the potential of *NewAthena* [17] and *Lynx* [18]. Yet with *Rubin* [19] and many other wide-field and time-domain oriented facilities in operation or anticipated, there is no planned X-ray mission that can provide comparably wide-field imaging spectroscopy, let alone at high spectral resolution.

*NewAthena* and *Lynx* will be transformative, but also narrow-field instruments (~[0.7,0.17] $deg^2$)[17, 18] in comparison to *Rubin* (9.6 $deg^2$) [19], and compared to localizations for LIGO, and for early spectroscopy of LISA inspirals (~10 $deg^2$ uncertainty at ~1 month pre-merger [20]).

A strategic complement to *NewAthena* and *Lynx* or any other realistic NASA X-ray flagship would be a **dedicated wide-field surveyor** *XESS* which simultaneously enables **spectroscopic and time domain capabilities** far in excess of what has been done by previous surveyors (e.g. *eROSITA*), and superior sensitivity and spectroscopic capability to what is possible for all-sky monitors or lobster optics.

*XESS* will provide a systematic, high-throughput surveyor and monitoring complement to X-ray flagships with superior sensitivity and spectroscopic capabilities but narrow (~10s of arcminutes) fields of view (FOVs). A partnership analogous to *Rubin-GMT* or *SPHEREx-JWST* is urgently needed for optimal *NewAthena*/*Lynx* targeting decisions.

*XESS* will conduct:

**1)** Deep, high angular/spectroscopic systematic surveys of the WHIM (Warm-Hot Intergalactic Medium) surrounding galaxies, and of large-scale cosmic structure.
**2)** Systematic surveys of AGN feedback and obscured AGN via high resolution spectroscopy (>3000 all-sky survey counts for ***all*** $L_X$~$10^{45}$ erg/s AGN at z<2).
**3)** Time domain spectroscopic monitoring of AGN for accretion state changes, quasi-periodic oscillations and ultrafast outflows (~daily at the North Ecliptic Pole).
**4)** High quality, low-interruption spectroscopic monitoring campaigns of compact objects in the galactic center and the Magellanic Clouds, including potentially constraints on the neutron star equation of state via phase-resolved absorption line gravitational redshifts.
**5)** Detection of rare, distant, X-ray bright transients and spectroscopic characterization (e.g. tidal disruption events [TDEs, >1000/year] and quasi-periodic eruptions [QPEs, >1/day]).
**6)** Efficient localization and spectroscopy of electromagnetic counterparts from GW transients, both post-merger LIGO events and pre-merger LISA events.
**7)** High spectral quality, high throughput uninterrupted monitoring of Milky Way stars for exoplanet occultations, coronal evolution and flares.
**8)** A complete spectroscopic census of Milky Way and Magellanic supernova remnants.
**9)** Complete resolved spectroscopic measurements of the local bubble, the hot circum-Galactic medium, Galactic outflows, shocks and the Fermi Bubbles, and potentially the solar wind charge exchange (SWCX).

*XESS* fills an essential gap in NASA's astrophysics capabilities and greatly expands those of all other facilities in X-rays and other bands in the 2030s astrophysics landscape.

## 2. Science:

| Science Objectives | Physical Parameters | Observables | Potential Challenges |
|---|---|---|---|
| *The nature of dark matter and dark energy*<br><br>*Mapping cosmic structure in clusters and filaments*<br><br>*Detect and map WHIM in both emission and absorption*<br><br>*Mechanisms for formation and evolution of BHs. The importance of hidden growth*<br><br>*Understanding AGN feedback physics.*<br><br>*Understanding stellar evolution and its end states, extreme gravity*<br><br>*Constraint the NS equation of state*<br><br>*Understand the history and evolution of our own Milky Way galaxy.* | *Distribution of baryons, correlation with structure. Temperature, ionization, density, entropy, pressure, metallicity, cooling times*<br><br>*Statistical census: contributions of shocks & reflection to AGN. Shock temperatures & integrated energies*<br><br>*BH masses, number density of obscured massive BHs across cosmic time. Outflow kinetic energy & evolution*<br><br>*Prevalence, environments & energetics of rare transients (GW events, tidal disruption events, exotic supernovae)*<br><br>*Supernova remnant spectroscopic census*<br><br>*Local BH/NS object compactness, gravity, geometry.* | *X-ray WHIM tracers at z~0: O VII, O VIII, Ne IX, N VII, C VI, Fe XVII, Mg XI*<br><br>*Neutral and hydrogen/helium-like transitions between Oxygen and Iron, z<6*<br><br>*Reflected Fe $K\alpha$ from AGN at z<6 for individual galaxies, stacked populations with optical, infrared priors.*<br><br>*Time domain monitoring: continuum & Fe line variability (ultrafast outflows:blueshifted Fe XXV, XXVI), reverberation mapping, quasi-periodic eruptions.*<br><br>*Localized X-rays from GW events (LIGO, LISA).*<br><br>*Spectral variability of Galactic and Magellanic compact objects.* | *Good sensitivity, (effective area $A_{eff}$, angular resolution), superior to eROSITA and XRISM, and large grasp (eROSITA-like or better).*<br><br>*Large field-of-view, matched to optical and GW facilities. Wide-field monitors lack sensitivity & resolution.*<br><br>*Wolter telescopes are massive, have required trade-offs between FOV & bandpass.*<br><br>*Good energy resolution (<10 eV). Cryogenics are essential, but how to keep weight & power down for a wide field? 0.05 K (microcalorimeter) is prohibitive for multiple modules to increase FOV, $A_{eff}$.*<br><br>*High photon throughput for bright source timing.* |

## 3. Instrument Description:

*XESS* dramatically improves upon *eROSITA*, in all respects: angular resolution, $A_{eff}$, FOV, energy resolution $\Delta E$, comparable bandpass (0.1-8.0 keV, with improved $A_{eff}$ at ~7 keV) and timing capability ($\Delta t$~few $\mu$s resolution, >1000 counts $s^{-1}$ $pixel^{-1}$ at nominal $\Delta E$).

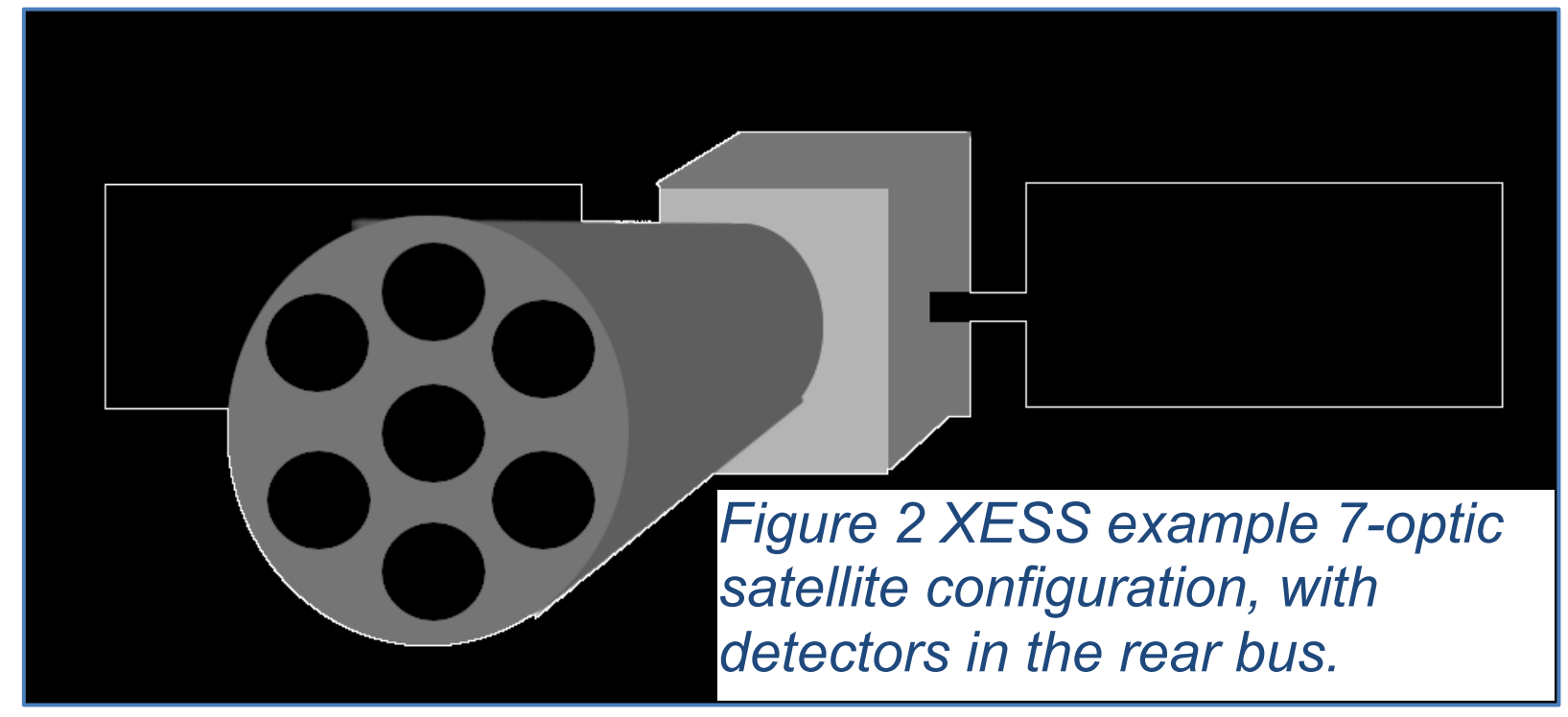

*Figure 2 XESS example 7-optic satellite configuration, with detectors in the rear bus.*

The base optics configuration leverages electroformed nickel replicated (ENR) Wolter-style optics (like *eROSITA*) for cost-effective maximization of grasp and FOV while retaining excellent $A_{eff}$ and angular resolution. Our strawman concept envisions seven modules, like *eROSITA*, but with important improvements using MSFC's ENR optics:

1) Retain a "fast" (D>1°) per-module FOV with large graze angles
2) Scale the per-module diameter to achieve *XMM-Newton* like $A_{eff}$ per module (~2000 $cm^2$, compare vs. the co-aligned *eROSITA* array) & ~5m focal length.
3) Use a polynomial prescription instead of pure Wolter I paraboloid-hyperbolid, to enable $\Delta\theta_{HPD}$~10" angular resolution and reduced vignetting across the field.
4) Use depth-graded multilayer coatings with low-density spacer layers to enable enhanced reflectivity out to ~7 keV at shallow graze, while using an iridium "cap" layer to retain high reflectivity below ~2 keV.
5) *Offset* aimpoints of six individual modules by ~1 FOV from the central module, such that the full etendue is $\Omega$ ~7 $deg^2$ with a D~3° FOV, comparable to the *Rubin* FOV and relevant GW uncertainties and therefore a match to their needs.

This configuration enables an extraordinary grasp, an order of magnitude greater than *eROSITA* and even above *Einstein Probe*, but with the effective area needed for both high resolution spectroscopy and variability studies beyond mere transient detection.

| Survey @ 1 keV | ~$A_{eff}$ ($cm^2$) | $\Omega$ ($deg^2$) | Grasp ($cm^2deg^2$) | $\Delta E$ (eV) | $\Delta\theta_{HPD}$ (arcsec) |
|---|---|---|---|---|---|
| ***XESS*** | **2000** | **7.000** | **14,000** | **10** | **10** |
| *eROSITA* [15] | 1300 | 1.000 | 1300 | 70 | 26 |
| *Einstein Probe* [21] | 3 | 3600.000 | 10,800 | 170 | 300 |
| *LEM* [22] | 2500 | 0.250 | 625 | 2 | 15 |
| *NewAthena X-IFU* [17] | 6000 | 0.004 | 21 | 4 | 9 |
| *Lynx LXM* [18] | 20000 | 0.007 | 139 | 3 | 0.5 |

While microcalorimeters would provide superior energy resolution, 10 eV is sufficient to detect and distinguish most relevant lines and ~500-3000 km/s shifts/broadening for many science cases. Meanwhile, Nyquist sampling of a 10” PSF over a 1° FOV will eventually require ~megapixel detectors. Cooling a 7-detector optical bench to a tight 0.05 K presents major obstacles in terms of SWaP-C (size, weight, power, cost) and limits mirror cost.

Dual Read-Out Imaging Detectors (DROIDs) using Superconducting Tunnel Junctions (STJs) have a long heritage of development for astrophysical use. They were considered for *XEUS* [23], and STJs are now mature high-count, μs-resolution beamline detectors [24,25]. Maturing STJs and DROIDs for X-ray astrophysics could enable ~10 eV resolution megapixel detectors with excellent cryogenic tolerance (“sloppy” 0.3 K, $He^3$ cooling, JFET readouts up to room temperature) to produce a replication-driven large-grasp X-ray mission, possibly aided by passive (*PRIMA*-like) optical bench cooling [26].

In principle, the mirror modules might be articulated to allow for full effective area on a narrower field of view. The practicality of such a capability will depend upon constraints from the detectors themselves, and associated cryocooling methods.

**4. Mission Implementation:**

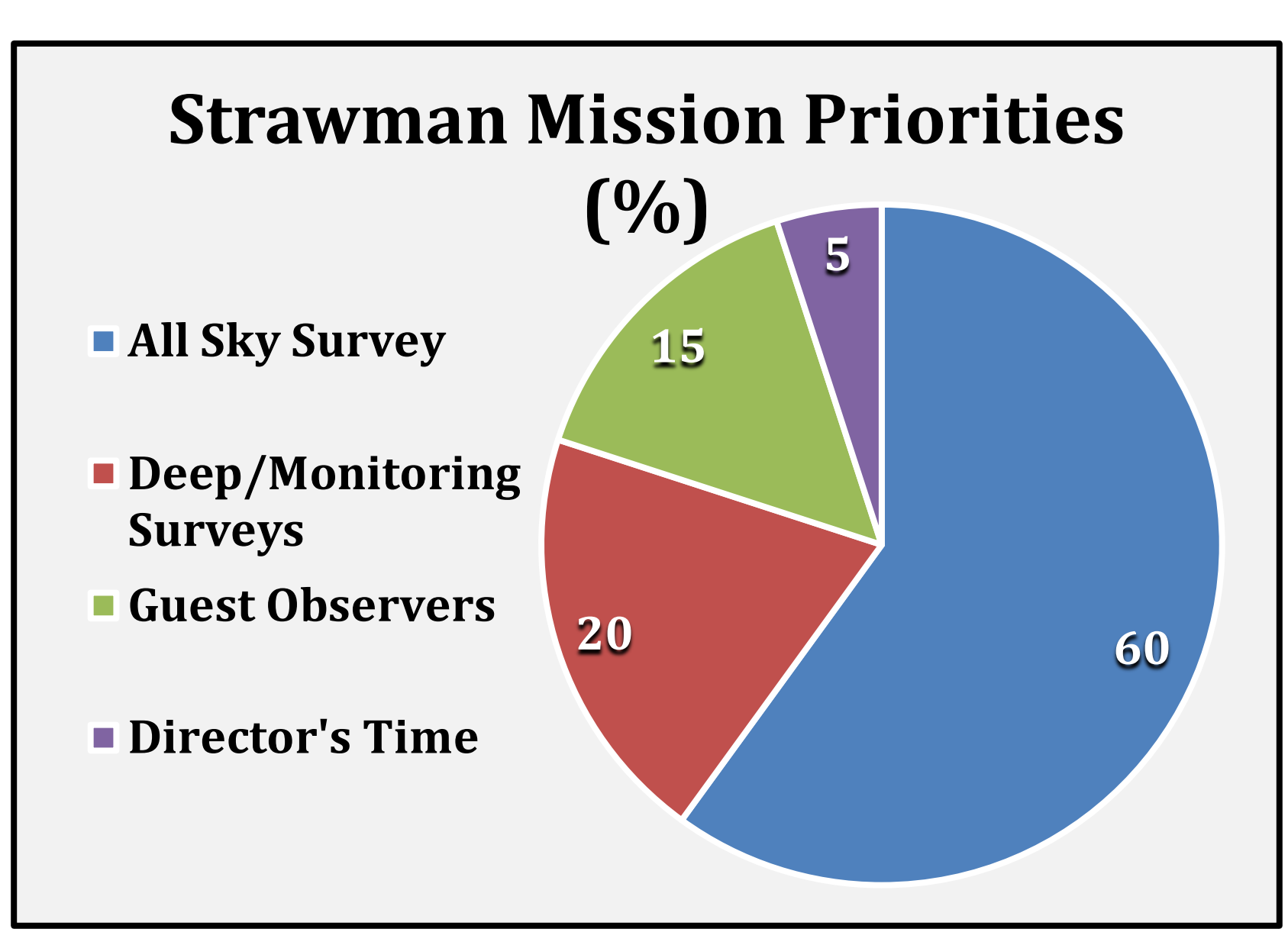

The basic configuration of *XESS* is a single spacecraft with optics, detectors and bus at L2 for modest background and maximum all-sky visibility. Since only the detectors must be cooled, a more flexible passive cooling scheme might permit greater visibility than *JWST* or *PRIMA*, but a full cryo chain to 0.3 K should be plausible for a <$2B mission if STJ readouts are in the warmest cooling stage. Fairing size and launch mass to L2 scaled from recent ENR mission mass/area ratios should be well within capabilities of a super-heavy lifter like Starship or New Glenn, with mass margin to aid fabrication & integration

costs (e.g. stiffer mirrors), and may even permit a more modest launch vehicle. Alternately, location at L1 might enable studies of the SWCX in the Earth's magnetosheath without loss to astrophysics.

The observing program would be similar to *Roman*, with most time devoted to surveys and a portion reserved for comprehensive guest observer programs. A strawman 5-year portfolio would be a 3-year all-sky survey, broken up into segments to incorporate all-sky time domain monitoring (much like *eROSITA*), 1 year dedicated to deep surveys and field monitoring (e.g. the Galactic center, the Magellanic Clouds, Andromeda, nearby clusters like Coma & Perseus), and 1 year for guest observers, e.g. Targets of Opportunity and Director's Discretionary Time. With high saturation rates for STJs in optimized regions, *XESS* can **simultaneously** monitor bright sources **and** survey adjacent fields. Given the nature of the facility and the value of both long-term monitoring and deep exposure, a longer mission would be preferred (with sustainability-optimized ground segment costs). Unlike *eROSITA,* all-sky results will promptly be available to the community.

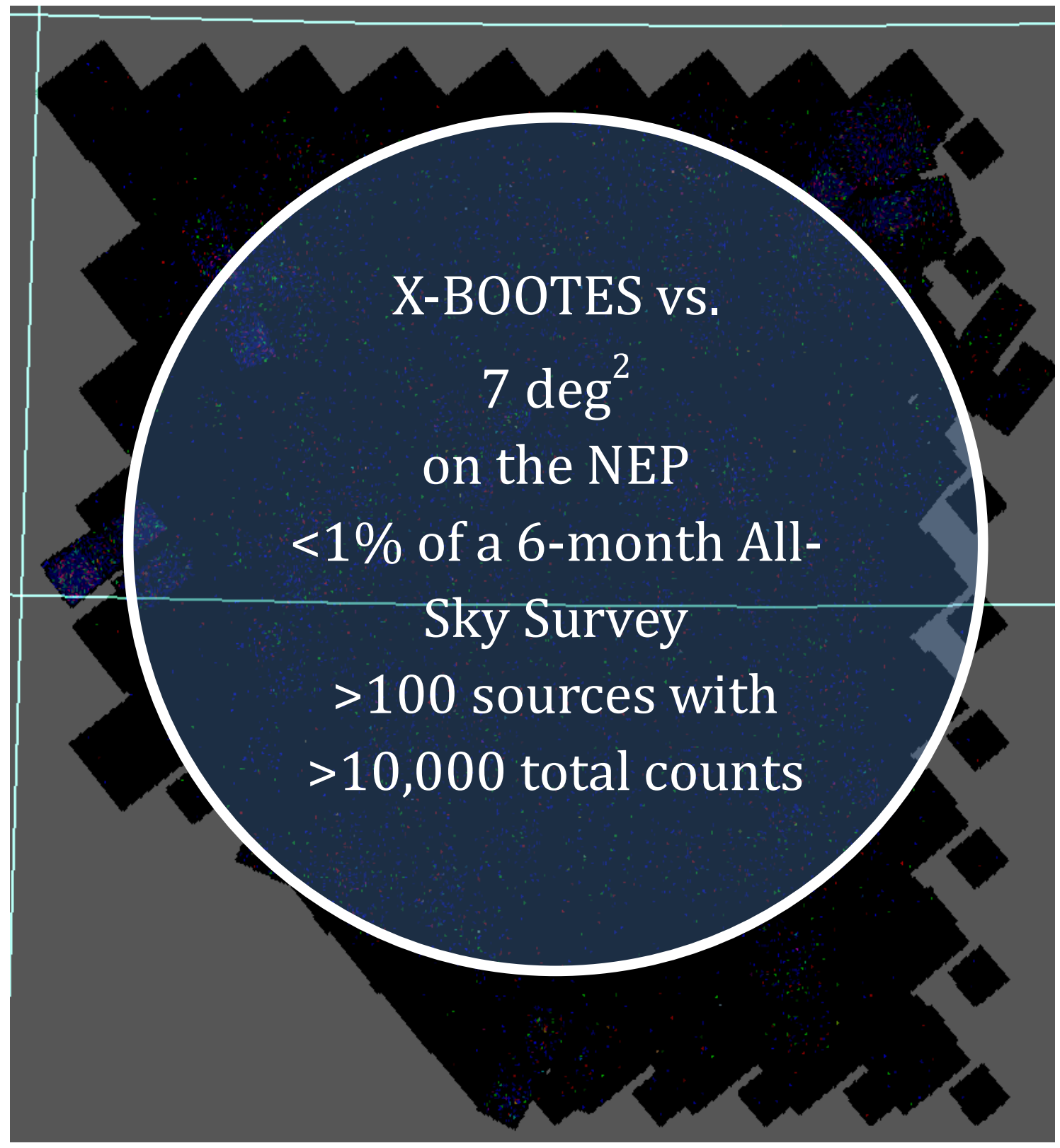

*Figure 3 ESA Sky, NASA, Chandra X-ray Observatory. Circle is an approximation of the Field-of-View, which will actually be 7 overlapping, vignetted ~1-deg circles.*

*X-BOOTES is Chandra's widest extragalactic survey.*

Although *XESS* is envisioned as a singular mission, the modular design implies conceptual flexibility. The largest hurdles will be producing a single, complete demonstration module with the optics, detectors and cryogenics. This module could be self-contained and scalable for up-scoping or descoping. Such adjustments might include a single multi-module spacecraft (adjusting either FOV or $A_{eff}$, depending upon relative aimpoints, and possibly incorporating a foldable design for an even larger mission, given the modest focal length).

In principle, heavy lifters like New Glenn and Starship can accommodate many more modules, with modules roughly co-aligned with either the rocket's length or (due to the short focal length) perpendicular to fit in a "pan pipes" configuration. Alternately, a smaller launch vehicle might be used with fewer modules (even one), such that the prototype enables industrial scaling of an iteratively built multi-satellite all-sky surveyor network as low as Low Earth Orbit, with a survey scanning mode pointed anti-Earth, and the potential for deeper observations via coordinated pointing.

An Artemis-driven version is possible, with twin units based at each pole (possibly passively cooled by craters) or in a lunar orbit to enable regolith and solar wind studies of the lunar surface. Space weather studies of the solar corona and charge exchange may be possible due to high count rate tolerance, but active regions would still saturate.

**Acknowledgements:**

Microsoft CoPilot AI was used to search literature and inform concept choices, to develop preliminary estimates of cost, and to augment optics trade-off analysis. No material was directly copied from AI outputs, no material was referred to without human review.